\documentclass[twocolumn,aps,prc,superscriptaddress,reprint]{revtex4-1}
\makeatletter

\newcommand{\Rmnum}[1]{\expandafter\@slowromancap\romannumeral #1@}
\makeatother
\usepackage{blindtext}
\usepackage{graphicx}
\usepackage{dcolumn}
\usepackage{bm}
\usepackage{booktabs}
\usepackage{color}
\usepackage{mathrsfs}
\usepackage{ulem}
\usepackage{verbatim}
\usepackage{indentfirst}
\usepackage{float}
\usepackage{subfigure}
\usepackage{mathtools,nccmath}
\usepackage[colorlinks,
           bookmarks=true,
           linkcolor=blue,
           urlcolor=blue, 
            anchorcolor=black,
            citecolor=blue
            ]{hyperref}
\usepackage{hypcap}
\usepackage{amsmath}

\renewcommand\sout{\bgroup \color{red} \ULdepth=-.5ex \ULset}

\newcommand{\commentout}[1]{}

\usepackage{epsfig}
\graphicspath{{plots/}}

\begin{document}

\preprint{APS/123-QED}

\title{Practical considerations for the effect of finite coverage on the azimuthal dependence of global spin alignment}

\author{Chao Zhang}%
\affiliation{Key Laboratory of Quark and Lepton Physics (MOE) and Institute
of Particle Physics, Central China Normal University, Wuhan 430079, China}

\author{Peng Yang}%
\affiliation{Key Laboratory of Quark and Lepton Physics (MOE) and Institute
of Particle Physics, Central China Normal University, Wuhan 430079, China}
\author{Feng Liu}%

\affiliation{Key Laboratory of Quark and Lepton Physics (MOE) and Institute
of Particle Physics, Central China Normal University, Wuhan 430079, China}
\author{Biao Tu}
\email{tubiao@hust.edu.cn}

\affiliation{%
 Cancer Center, Union Hospital, Tongji Medical College, Huazhong University of Science and Technology, Wuhan 430022, China
}%

\date{\today}

\begin{abstract}
The global spin alignment of vector mesons is a powerful probe to study the vorticity field of the system produced in non-central relativistic heavy-ion collisions. However, the experimental correction of the global spin alignment observable remains to be a long standing issue. In this study, we propose a method to correct for the effect of finite pseudo-rapidity coverage when extracting the azimuthal dependence of spin alignment parameter $\rho_{00}$. The potentially existing intrinsic spin alignment effects are taken into consideration together with the acceptance correction. The method presented in this paper is verified in a Monte-Carlo simulation, it allows the measurements of azimuthal dependence of vector mesons global spin alignment to be conducted properly and accurately.

\end{abstract}

\maketitle


\section{\label{sec:intro}Introduction}
In non-central relativistic heavy-ion collisions, the two colliding nuclei travel at close-to-light speed, generating large initial orbital angular momentum. A considerable fraction of such orbital angular momentum is deposited in the system, via the process of spin-orbital coupling, partially reveals itself in the form of global polarization of produced particles at the final stage~\cite{Liang:2004ph,Liang:2004xn,Betz:2007kg,Gao:2007bc}. This effect can result in a polarization for the quarks, and it can be identified with the hyperon polarization and vector meson spin alignment in the experiments. Recently STAR collaboration reported a finite global $\Lambda$ hyperon polarization, indicating that a most vortical fluid, yet with least viscosity, has been created at RHIC~\cite{STAR:2017ckg}. The comprehensive studies about the global polarization of hyperons and spin alignment of vector mesons will provide us key information about the vorticity field~\cite{Becattini:2007sr,Pang:2016igs,Jiang:2016woz,Becattini:2015ska} created in and the dynamical information about the hot and dense matter. Therefore the topic is gaining increasing interests in the field of both theory~\cite{Zhang:2019xya, Huang:2011ru, Florkowski:2019qdp,Xia:2020tyd,Li:2017slc} and experiments~\cite{ALICE:2019aid,ALICE:2019onw,Singh:2018uad,Singha:2020qns,STAR:2018gyt,STAR:2008lcm}.

The initial twist generated from the spectator nucleons leads to maximized vorticity along the reaction plane(RP) direction. Since the system produced in the non-central relativistic heavy-ion collisions has low viscosity, the vorticity may not propagate efficiently from the in-plane direction to the out-plane direction, which results in stronger in-plane global polarization. In Ref.~\cite{Niida:2018hfw}, the STAR collaboration observed a larger polarization in in-plane direction for $\Lambda$ and $\bar{\Lambda}$ hyperons in Au+Au collisions at $\sqrt{s_{NN}}$ = 200~GeV. The trend can be explained by the hydrodynamics including the contribution of thermal vorticity and shear viscosity~\cite{Fu:2021pok,Sun:2021nsg}. A similar study with ``isothermal local equilibrium" scenario~\cite{Becattini:2021iol} can also explain the azimuthal angle dependence. The azimuthal dependence of spin alignment may also carry information of the system evolution. However, the azimuthal dependence of vector meson spin alignment is not fully explored in previous studies up to now. 

The spin alignment of spin-1 vector mesons is described by a 3$\times$3 Hermitian spin-density matrix $\rho$ with unit trace~\cite{Schilling:1969um}. Since the diagonal matrix elements $\rho_{11}$ and  $\rho_{-1-1}$ are degenerated, the net spin alignment is detected from the deviation of the $\rho_{00}$ parameters from 1/3. The $\phi$ spin alignment parameter ($\rho_{00}$) can be determined by measuring the angular distribution of the $\phi$ decay products with respect to the orbital angular momentum of system $\textbf{L}$:

\begin{eqnarray}
\frac{dN}{d(\mathrm{cos}\theta^{\star})} \propto (1-\rho_{00}) + (3 \rho_{00} -1) \mathrm{cos}^2\theta^{\star},
\label{eq:rho_00}
\end{eqnarray}

where $\theta^{\star}$ is the angle between $\textbf{L}$ and the momentum of the daughter kaon in the parent $\phi$ meson's rest frame. $\rho_{00}$ can be extracted from the angular distribution of Eq.~\ref{eq:rho_00} in the experiments, a deviation of the $\rho_{00}$ from 1/3 signals the spin alignment of vector mesons.

In Ref.~\cite{Lan:2017nye}, by toy model simulation, it is found that finite coverage of pseudo-rapidity can lead to an artificial enhancement of the global spin alignment signal. and its correction has been investigated in Ref.~\cite{Tang:2018qtu}. Therefore, it is crucial to study whether the finite coverage would lead to an artificial azimuthal dependence of the spin alignment $\rho_{00}$. In this article, we aim to correct for the effect of finite coverage on the azimuthal dependence of vector mesons spin alignment. The rest of the paper is organized as follows: The effects of finite coverage of pseudo-rapidity on the azimuthal dependence of extracted spin alignment parameter $\rho_{00}$ are discussed in Sec.~\ref{sec:effonphi}. We then propose a method to correct for these effects in Sec.~\ref{sec:corr}. After we discuss the set up of a toy model and results in Sec.~\ref{sec:simu} and Sec.~\ref{sec:res}, a summary is then given in Sec.~\ref{sec:sum}.

\section{\label{sec:effonphi}Effect on the azimuthal dependence of $\phi$ spin alignment}

The measurement of the spin alignment of vector mesons, e.g., $\phi$ meson, is equivalent to the study of 3-dimensional momentum distribution of the decay daughters in their parent rest frame. Therefore, it is desirable to conduct the measurement in full phase space. However the phase space coverage of the detectors is always limited, thus introducing bias to the extracted $\rho_{00}$ values of vector mesons. The effects induced by the finite $\eta$ coverage, and the potentially existing intrinsic spin alignment of $\phi$ meson can both lead to deviation of the extracted $\rho_{00}$ from 1/3. Here we discuss these effects and propose a method to correct for them.

\begin{figure}[!htb]
 \begin{center}
   \includegraphics[scale=0.30]{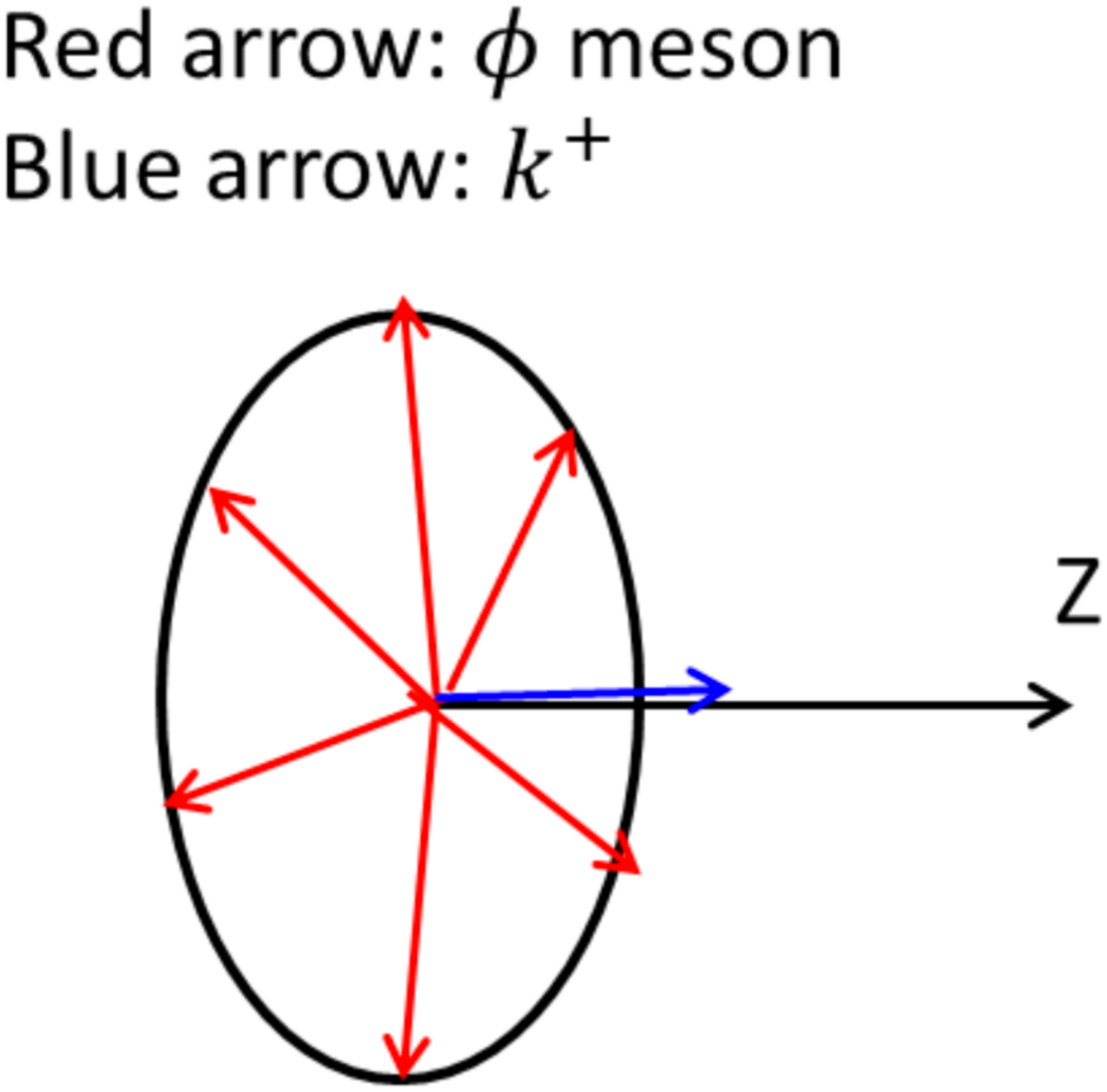}
        \caption{ The effect of finite pseudo-rapidity  coverage on the $\phi$ meson spin alignment viewed from the local helix frame.}
   \label{fig:1}
  \end{center}
 \end{figure}
\subsection{\label{sec:accept}Pseudo spin alignment}

\subsubsection{\label{sec:global}Pseudo global spin alignment}
The finite acceptance of the detector constrains the phase space of the daughter particles, and it can lead to an artificial global spin alignment of the $\phi$ meson. This is due to the narrow $\eta$ cut on the $K^{+}$ tends to exclude those decayed $\phi$ meson with $K^{+}$ at $\theta^{\star} \sim90^{\circ}$, e.g., as shown in Fig.~\ref{fig:1}, assuming the momentum of the decayed $\phi$ meson is along the transverse plane (shown as red arrows), and the decayed daughter $K^{+}$ with a momentum perpendicular to $\phi$ meson(shown as blue arrow). The finite $\eta$ cut excludes these cases which can lead to a global spin alignment of the $\phi$ mesons. Note that this effect has been pointed out in Ref.~\cite{Lan:2017nye}, and a method to correct for this effect has been proposed in Ref.~\cite{Tang:2018qtu}. 
\begin{figure}[!htb]
 \begin{center}
   \includegraphics[scale=0.3]{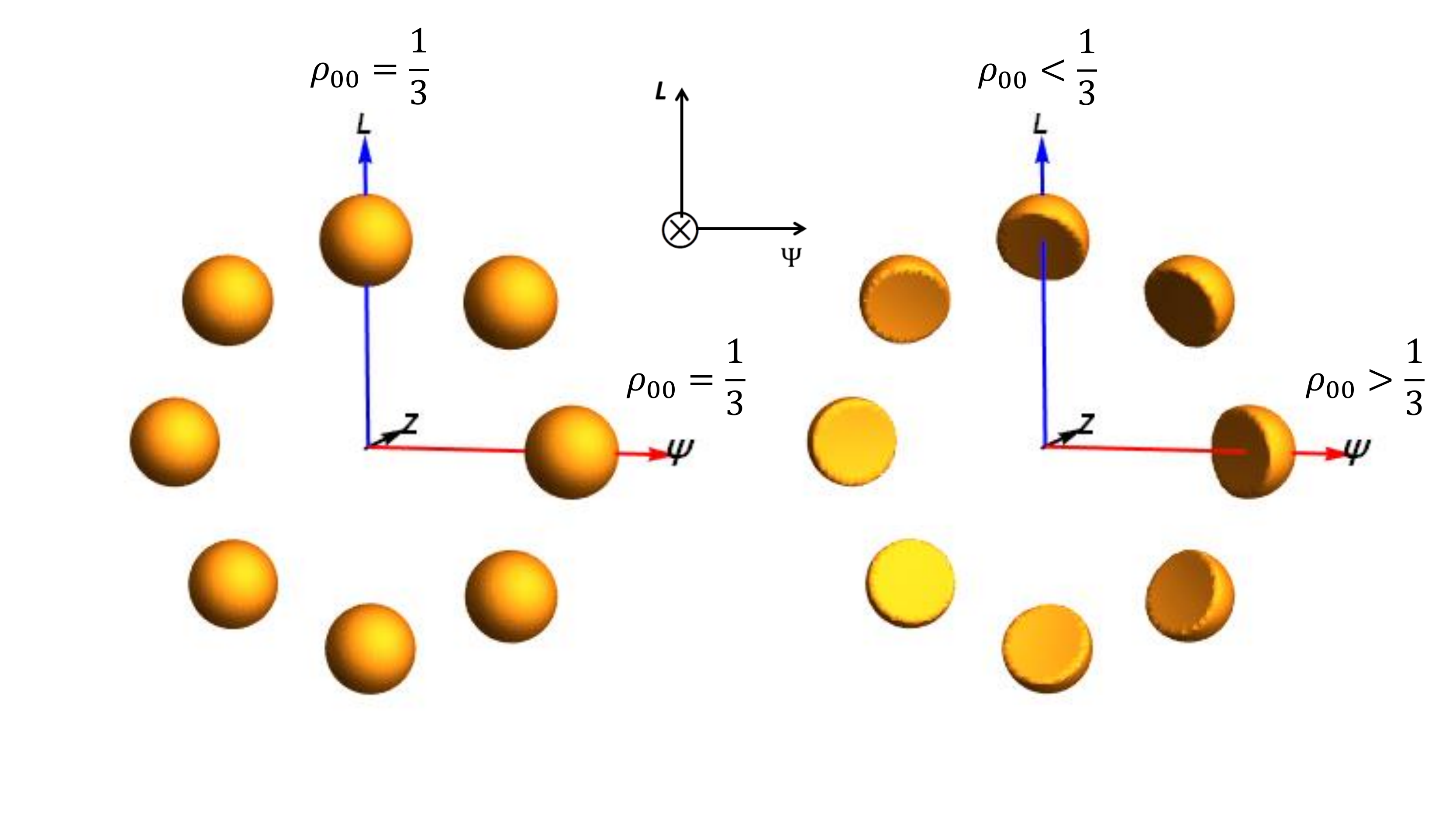}
        \caption{ The azimuthal dependence of effects from finite pseudo-rapidity coverage on the $\phi$ meson spin alignment viewed from the global frame. }
   \label{fig:2}
  \end{center}
 \end{figure}

\subsubsection{\label{sec:local}Pseudo local spin alignment}
The finite pseudo-rapidity acceptance also leads to the biased azimuthal angle dependence on the $\phi$ meson spin alignment measurement. To study these effects, one needs to study the azimuthal angle dependence of spin alignment parameter $\rho_{00}$. As shown in Fig.~\ref{fig:2}: Assuming there are no spin alignment of the $\phi$ meson: e.g. $\rho_{00}=1/3$, the 3-dimension momentum distribution of the $K^{+}$ in the $\phi$ meson rest frame is in a ball-like shape. This is shown on the left side of Fig.~\ref{fig:2} when there is no $\eta$ cut on the phase space. The ball-like shape changed to ellipsoid with the major axis along the $\phi$ meson if switching to the lab frame. However, each ellipsoid is dug with a cone along the $Z$ axis once there exists a finite $\eta$ cut. Therefore, as shown in the right hand side, the 3-dimension momentum distribution of the $K^{+}$ in the $ \phi $ meson rest frame deviates from the ball-like shape which induced a deviation of $\rho_{00}$ from 1/3. In addition, the effects from the finite $\eta$ cut on the $\phi$ meson spin alignment depend on the azimuthal angle when viewed from the local frame. From Fig.~\ref{fig:2}, the $\rho_{00,acc}$ is greater than 1/3 along the RP direction and less than 1/3 along the direction perpendicular to RP.

The effects induced by the finite pseudo-rapidity coverage on the $\phi$ meson spin alignment have been discussed in sec.~\ref{sec:global} and ~\ref{sec:local} from two perspectives, i.e., viewed from the global frame and local frame, respectively. It is known that the global and local spin alignment are intra-connected, which means the change in one will result in a change in the other~\cite{ALICE:2019aid}. Therefore, in this work the acceptance constrained spin alignment is denoted as $\rho_{00,acc}$.

\subsection{\label{sec:intrinsic}Intrinsic spin alignment}
\begin{figure}[!htb]
 \begin{center}
   \includegraphics[scale=0.45]{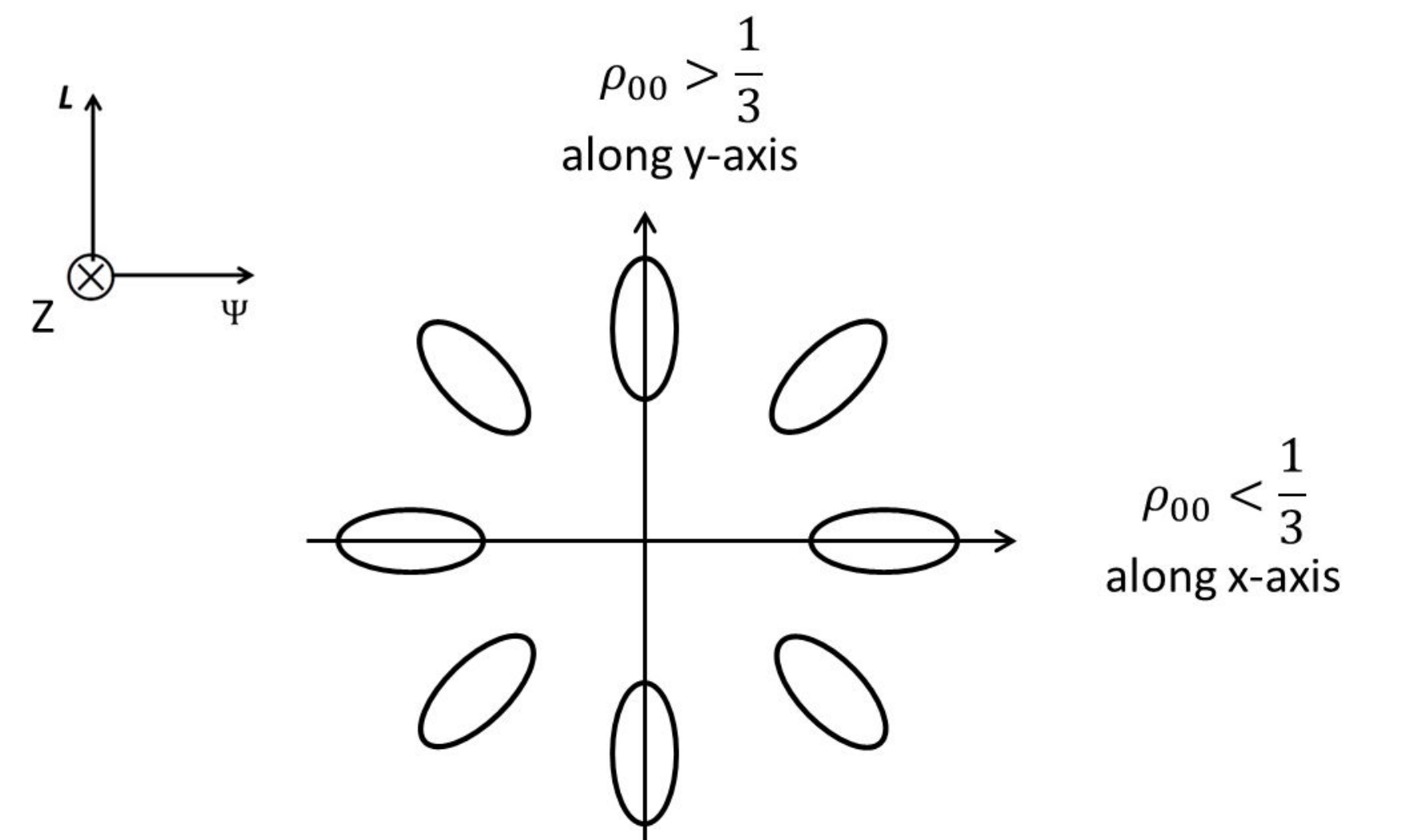}
        \caption{ The azimuthal dependence of effects from the potentially exist intrinsic spin alignment in the local helix frame on the $\phi$ meson spin alignment.  }
   \label{fig:3}
  \end{center}
 \end{figure}

The effects of possible intrinsic spin alignment in particle's helix frame can be taken into consideration, even in the lack of measured value of intrinsic spin alignment for the $\phi$ meson. Assuming the $\phi$ meson has a intrinsic spin alignment of $\rho_{00}$ greater than 1/3 in the helix frame, the extracted $\rho^{HX}_{00,in}$ from the angular distribution of $K^{+}$ with respect to the orbital angular momentum of system $\textbf{L}$ should depend on the azimuthal angle.
As shown in Fig.~\ref{fig:3}, the extracted $\rho^{HX}_{00,in}$ is less than 1/3 along the RP direction and $\rho^{HX}_{00,in}$ is greater than 1/3 along the direction perpendicular to RP.
\section{\label{sec:corr}Correction method}
The angular momentum $\textbf{L}$ of the system is the quantization axis when measuring the $\rho_{00,acc}$ and $\rho^{HX}_{00,in}$. However will those effects exist if one randomly rotates the quantization axis in the 3-dimensional space (3D rotation)? Note that the random rotation of the quantization axis is equivalent to random rotation of the particles' momentum. The effect of 3D rotation we discussed in this section is for the extracted $\rho_{00}$ with a given azimuthal angle. As the 3D rotation will not disturb the integrated $\rho_{00}$ values, but only change the value of $\rho_{00}$ for a given azimuthal angle. The effects on $\rho_{00,acc}$ won't be vanished due to the particle selection within the finite pseudo-rapidity window, but the value of $\rho_{00,acc}$ will be changed. Therefore, denote as $\rho_{00,acc,3D}$, they still have azimuthal angle dependence. As for the intrinsic spin alignment shown in Fig.~\ref{fig:3}, the 3D rotation will not affect the $\rho^{HX}_{00,in}$, they also have the azimuthal angle dependence.

We use a Monte-Carlo simulation to correct the effects of the finite $\eta$ cut and the intrinsic spin alignment effects on the measurement of the real spin alignment of $\phi$ meson $\rho_{00,real}$. In the experiment data, it contains the combined effects of the following:
\begin{eqnarray}
\begin{aligned}
\rho_{00,acc}~,~\rho^{HX}_{00,in}~,~\rho_{00,real}
\end{aligned}
\end{eqnarray}

After the 3D rotation was applied to the experiment data, the $\rho_{00,real}$ vanished, and it contains the combine effects of:
\begin{eqnarray}
\begin{aligned}
\rho_{00,acc,3D}~,~\rho^{HX}_{00,in}
\end{aligned}
\end{eqnarray}

In the simulation data, the $\rho_{00,real}=1/3$ and there are no intrinsic spin alignment for the $\phi$ meson, therefore it only contains:
\begin{eqnarray}
\begin{aligned}
\rho_{00,acc}
\end{aligned}
\end{eqnarray}

After the 3D rotation was applied, it contains:
\begin{eqnarray}
\begin{aligned}
\rho_{00,acc,3D}
\end{aligned}
\end{eqnarray}

According to the discussion above, we can extract the additional spin alignment parameters $\rho_{00}$ from different effects with the following steps:

As shown in Eq.~\ref{eq:6}, We first extract the $\rho_{00,acc,3D}$ by fitting the angular distribution of the 3D rotated simulation data.
\begin{eqnarray}
\begin{aligned}
 \frac{dN_{Simu\_3D}}{d(\mathrm{cos}\theta^{\star})}  & \propto (1-\rho_{00,acc,3D}) + (3 \rho_{00,acc,3D} -1) \mathrm{cos}^2\theta^{\star},
  \label{eq:6}
\end{aligned}
\end{eqnarray}

The intrinsic spin alignment factors $\rho^{HX}_{00,in}
$ can be extracted from the angular distribution of the 3D rotated experiment data according to Eq.~\ref{eq:7},

\begin{eqnarray}
\begin{aligned}
\frac{dN_{Data\_3D}}{d(\mathrm{cos}\theta^{\star})} & \propto [(1-\rho^{HX}_{00,in}) + (3 \rho^{HX}_{00,in} -1) \mathrm{cos}^2\theta^{\star}]  \\
&\times [(1-\rho_{00,acc,3D}) + (3 \rho_{00,acc,3D} -1) \mathrm{cos}^2\theta^{\star}],
\label{eq:7}
\end{aligned}
\end{eqnarray}

After we extracted the $\rho_{00,acc}$ from the angular distribution of the simulation data,
\begin{eqnarray}
\begin{aligned}
 \frac{dN_{Simu}}{d(\mathrm{cos}\theta^{\star})}
 \propto (1-\rho_{00,acc}) + (3 \rho_{00,acc} -1) \mathrm{cos}^2\theta^{\star},
\end{aligned}
\end{eqnarray}

The real spin alignment parameter $\rho_{00,real}$ can be extracted by fitting the angular distribution of experiment data:

\begin{eqnarray}
\begin{aligned}
\frac{dN_{Data}}{d(\mathrm{cos}\theta^{\star})} & \propto  [(1-\rho_{00,acc}) + (3 \rho_{00,acc} -1) \mathrm{cos}^2\theta^{\star} ]\\
&\times [(1-\rho^{HX}_{00,in}) + (3 \rho^{HX}_{00,in} -1) \mathrm{cos}^2\theta^{\star}] \\
&\times [(1-\rho_{00,real}) + (3 \rho_{00,real} -1) \mathrm{cos}^2\theta^{\star}],
\end{aligned}
\end{eqnarray}

The above method to correct for the effects of the finite acceptance and intrinsic spin alignment on the measurement of azimuthal dependence of global spin alignment has been examined by a toy model simulation.

\section{\label{sec:simu}Setup of toy model simulation}

We construct a toy model which samples the phi meson distribution based on the $p_T$ spectrum, $dN/dy$ distribution and elliptic flow $v_2$~\cite{STAR:2007mum} measured by the STAR experiment. PYTHIA6 model was then used to simulate the $\phi \to K^{+} + K^{-}$ decay process~\cite{Sjostrand:2014zea}. The reaction plane angle $\hat{\Psi}$ and $\hat{p}_{beam}$ are set to(1,0,0), (0,0,1), respectively. Therefore according to Ref.~\cite{STAR:2007ccu} the angular momentum of the system $\textbf{L}$ is set to (0,-1,0). Different $\rho_{00,real}$ parameters were taken as input at different $\Phi-\Psi$ bins. In the $\phi$ meson helicity frame where the $\theta^{*}$ is the angle between the $K^{+}$ momentum in the $\phi$ meson rest frame with respect to the $\phi$ meson momentum, the intrinsic spin alignment parameter was set according to Eq.~\ref{eq:rho_00}. One can measure the $K^{+}$ angular distribution with respect to $\textbf{L}$ at different $\Phi-\Psi$ bins after the finite $\eta$ cut was applied, and the combined $\rho_{00}$ from $\rho_{00,acc}~,~\rho^{HX}_{00,in}$ and $\rho_{00,real}$ for different $\Phi-\Psi$ bins can then be extracted. The combined $\rho_{00,acc,3D}$ and $\rho^{HX}_{00,in}$ can be extracted after we apply the 3D rotation to the quantization axis in the above system, note that the $\hat{p}_{beam}$ and the reaction plane angle $\hat{\Psi}$ are changed in 3D rotation. The above event samples are denoted as $``Data"$ and $``Data\_3D"$.

The above processes were repeated with the real spin alignment parameters $\rho_{00,real}$ and the intrinsic spin alignment parameters were both set to 1/3. One can extract the $\rho_{00,acc}$ or $\rho_{00,acc,3D}$ at different $\Phi-\Psi$ bins depending on whether the 3D rotation is applied. The event samples above are denoted as $``Simu"$ or $``Simu\_3D"$.

According to the above processes, the $\Phi-\Psi$ range:$(0-\pi/2)$ was divided into four equivalent bins. The input spin alignment parameters $\rho_{00,input}$ at different $\Phi-\Psi$ bins were set to (0.38,0.36,0.34,0.32), and the intrinsic spin alignment parameter was set to 0.35. We can now extract the spin alignment parameters $\rho_{00}$ for different effects according to Sec.~\ref{sec:corr}.

\section{\label{sec:res}Results and discussions}
\begin{figure}[!htb]
 \begin{center}
   \includegraphics[scale=0.45]{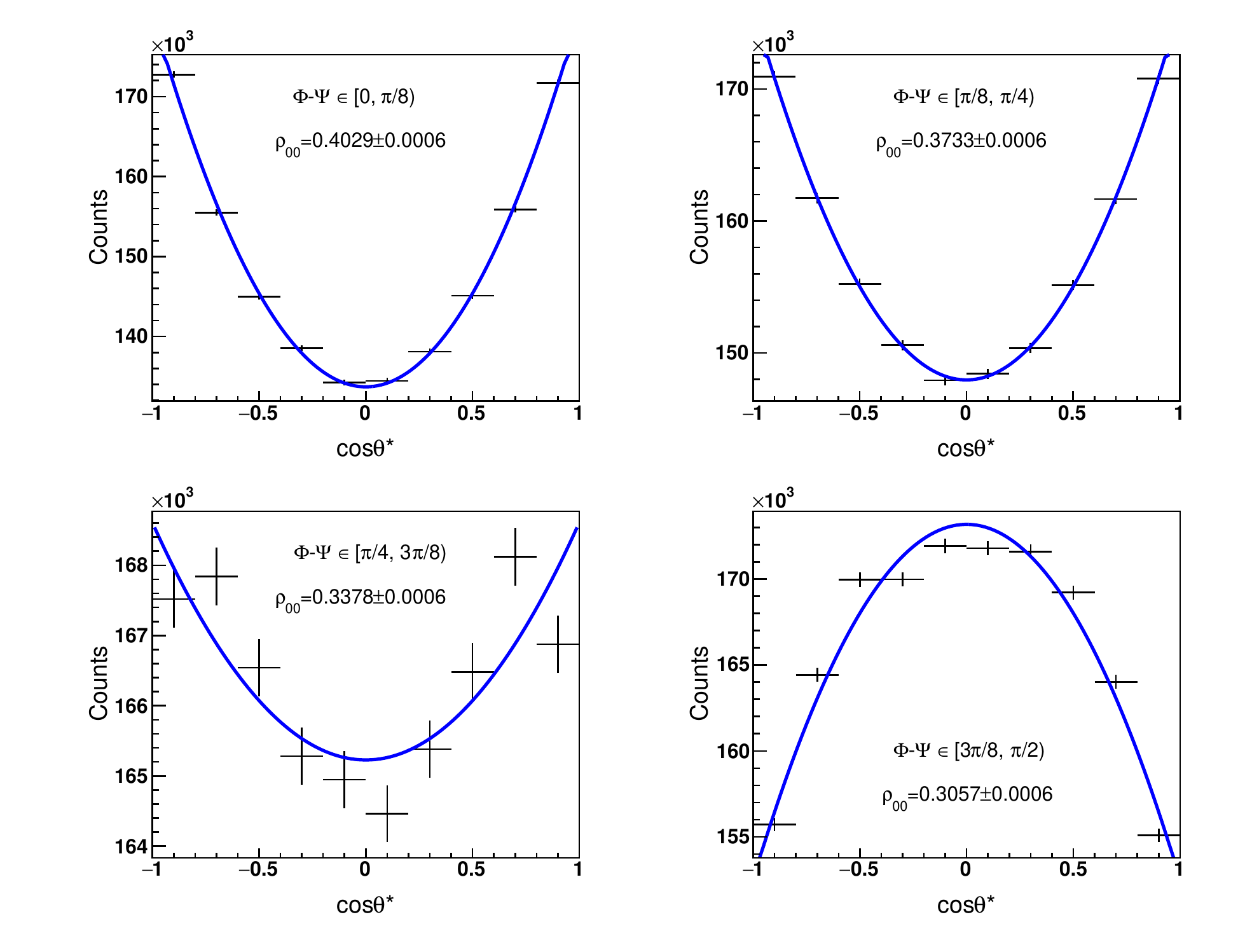}
        \caption{  The $cos\theta^{*}$ distribution for different $\Phi-\Psi$ bins from the ``Data" event sample, the Lines are the fitted function from Eq.~\ref{eq:rho_00}.  }
   \label{fig:4}
  \end{center}
 \end{figure}

Fig.~\ref{fig:4} shows the $cos(\theta^{*})$ distribution from the $``Data"$ event sample in different $\Phi-\Psi$ bins. The blue lines are fitted from Eq.~\ref{eq:rho_00} and the extracted $\rho_{00}$ values are also shown in the figure. The extracted spin alignment parameters $\rho_{00}$ show clearly deviation from our input $\rho_{00,input}(0.38,0.36,0.34,0.32)$ for all $\Phi-\Psi$ bins. Note that the values of the extracted $\rho_{00}$ parameters shown in Fig.~\ref{fig:4} are affected by both the finite pseudo-rapidity coverage $\rho_{00,acc}$ and the intrinsic spin alignment $\rho^{HX}_{00,in}$ of the $\phi$ meson.

%
 
\begin{figure}[!htb]
 \begin{center}
   \includegraphics[scale=0.4]{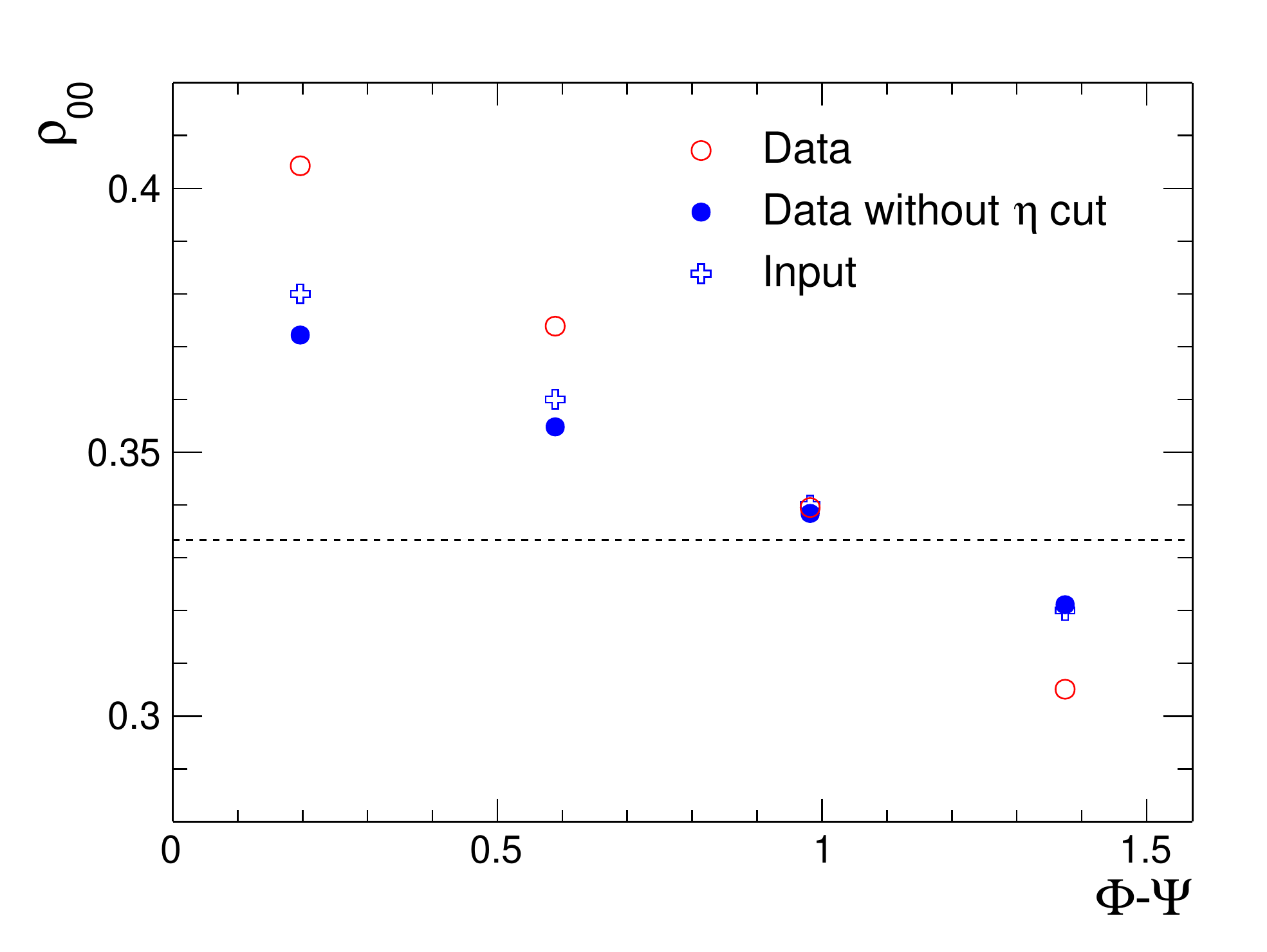}
        \caption{ The extracted $\phi$ meson spin alignment parameters $\rho_{00}$ as a function of $\Phi-\Psi$ from different Data sample with or without finite $\eta$ coverage in comparison with $\rho_{00,input}$.  }
   \label{fig:5}
  \end{center}
 \end{figure}
Fig.~\ref{fig:5} shows the extracted spin alignment parameters $\rho_{00}$ as function of $\Phi-\Psi$ from the event sample ``Data"  in comparison with $\rho_{00,input}$. The results labeled as ``Data without $\eta$ cut" represent the extracted $\rho_{00}$ parameters from the event sample "Data" with no finite pseudo-rapidity cut. The extracted  $\rho_{00}$ from the event sample ``Data", which affected by both finite pseudo-rapidity coverage $\rho_{00,acc}$ and intrinsic spin alignment $\rho^{HX}_{00,in}$, show strong deviation from the $\rho_{00,input}$. This is mostly due to the finite pseudo-rapidity coverage effects, which can be seen from the results of the event sample ``Data without $\eta$ cut".  The intrinsic spin alignment of $\phi$ meson also have effects on the azimuthal dependence of extracted spin alignment parameters $\rho_{00}$. These results are a prove of the importance of the correction for the finite acceptance and the intrinsic spin alignment effects when measuring the azimuthal dependence of $\phi$ meson global spin alignment.
\begin{figure}[!htb]
 \begin{center}
   \includegraphics[scale=0.4]{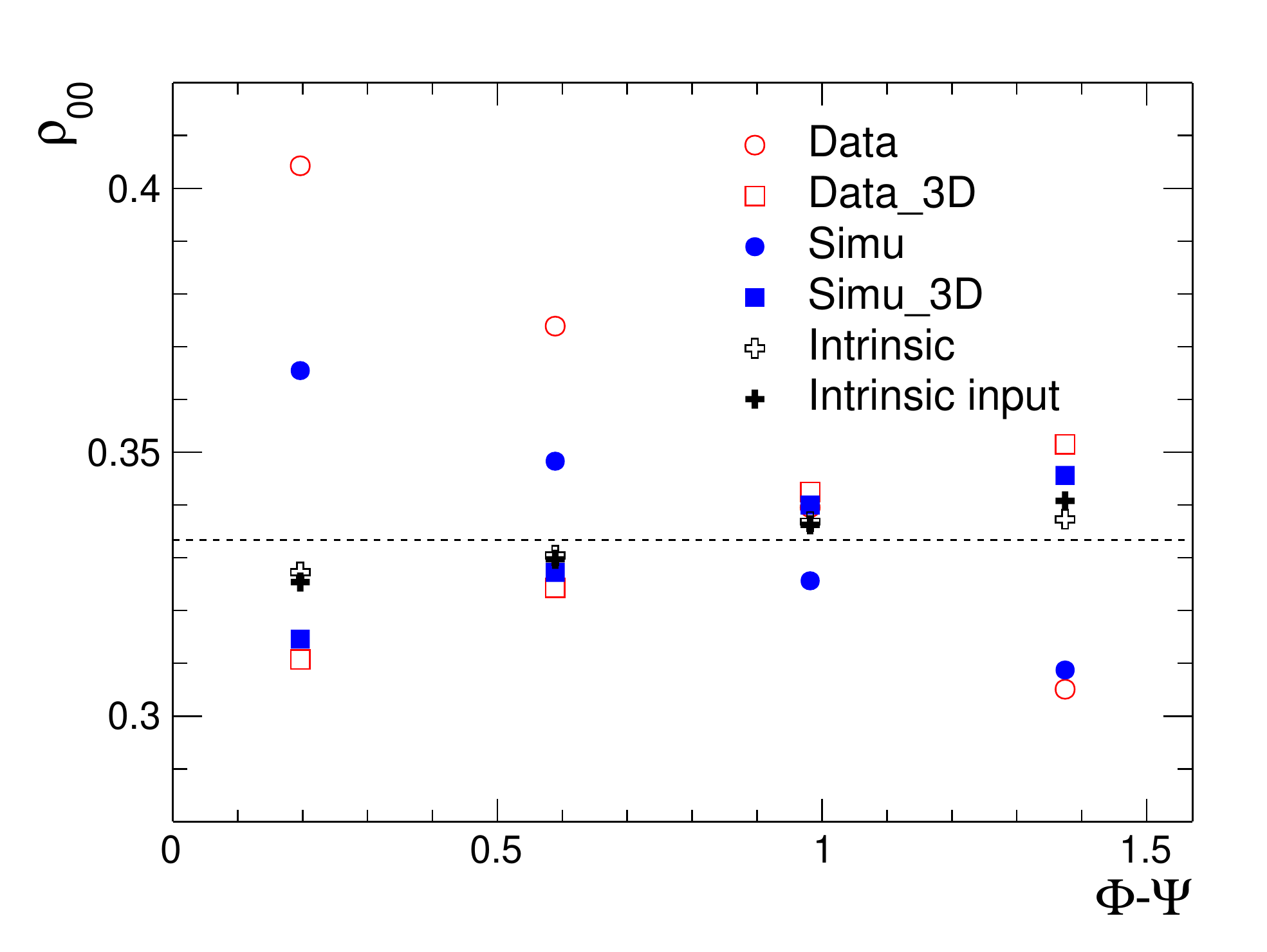}
        \caption{ The extracted $\rho^{HX}_{00,in}$ based on Eq.~\ref{eq:7}, $\rho^{HX}_{00,in}$ from input and extracted $\rho_{00}$ from event sample $``Data", ``Data\_3D",  ``Simu" and ``Simu\_3D"$ as a function of $\Phi-\Psi$.}
   \label{fig:6}
  \end{center}
 \end{figure}
 \begin{figure}[!htb]
 \begin{center}
   \includegraphics[scale=0.4]{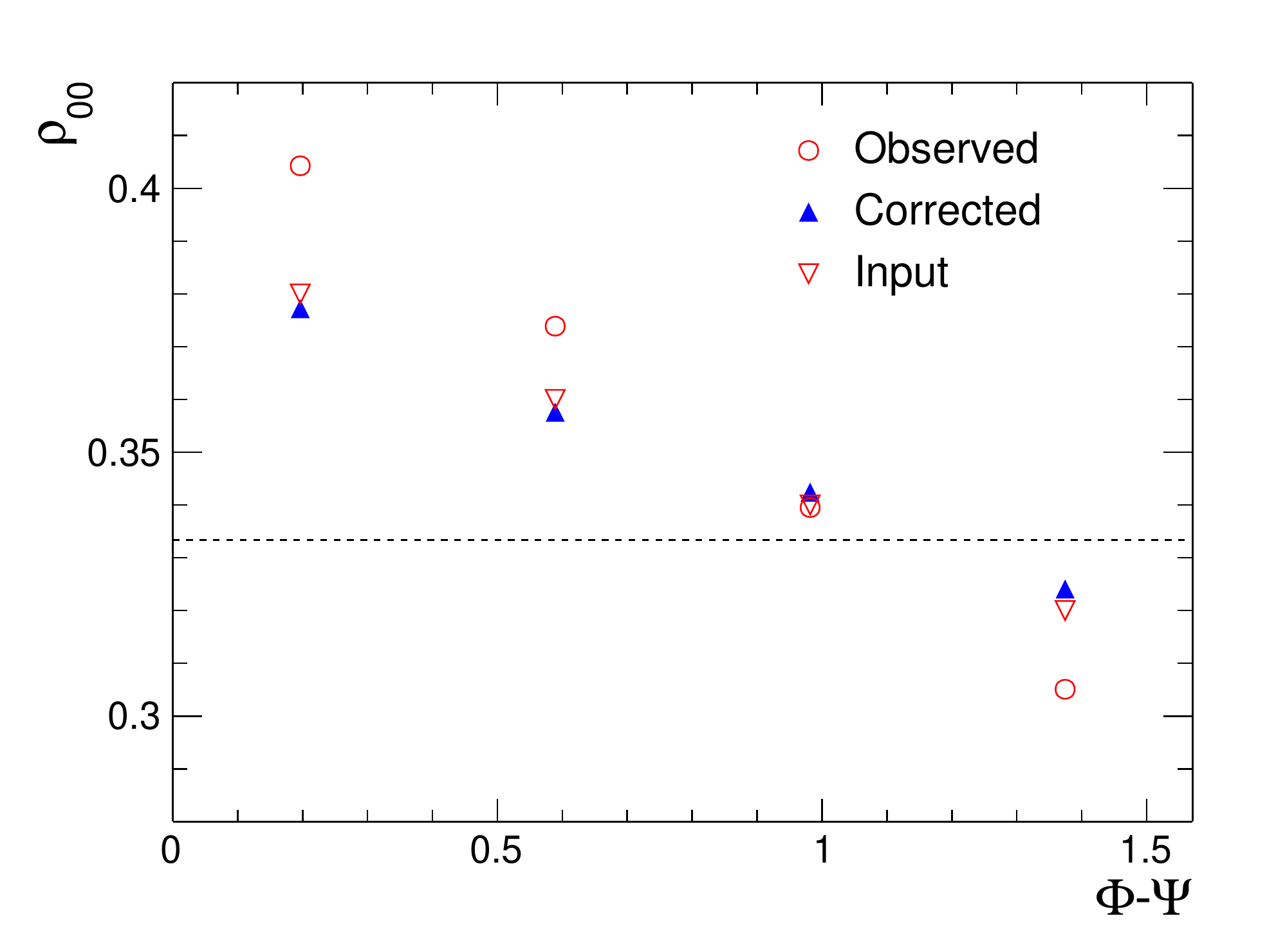}
        \caption{ Input, corrected and observed $\rho_{00}$ as a function of $\Phi-\Psi$. }
   \label{fig:7}
  \end{center}
 \end{figure}
 

Fig.~\ref{fig:6} shows the extracted $\rho_{00}$ values as function of $\Phi-\Psi$ for all four event samples mentioned in Sec.~\ref{sec:simu}. From the results, the extracted $\rho_{00}$ values from the event sample $``Data"$ show a stronger deviation to 1/3 comparing to the $\rho_{00,input}$, which were also shown in Fig.~\ref{fig:5}. After we apply the 3D rotation, as shown in the results of $``Data\_3D"$, the azimuthal dependence of extracted $\rho_{00}$ changed. For the event sample $``Simu"$ and $``Simu\_3D$'' where the $\rho_{00,real}$ and intrinsic spin alignment parameter $\rho_{00,in}^{HX}$ were set to 1/3, a deviation of the extracted $\rho_{00}$ from 1/3 have been observed for all four $\Phi-\Psi$ bins. Note that one can extract the $\rho^{HX}_{00,in}$ if they combine the results of $``Data\_3D"$ and $``Simu\_3D"$. Denoted as ``intrinsic' and ``intrinsic input'', respectively, the extracted $\rho^{HX}_{00,in}$ from Eq.~\ref{eq:7} and the $\rho^{HX}_{00,in}$ input are also shown in Fig.~\ref{fig:6}. Note that the input intrinsic spin alignment parameter was set to 0.35 in the $\phi$ meson helix frame, and the intrinsic input parameters are the extracted global $\rho_{00}$ caused by the intrinsic $\rho_{00}$ parameter in the helix frame. While there are small discrepancy, the intrinsic $\rho_{00}$ parameters extracted from Eq.~\ref{eq:7} are consistent with the intrinsic input parameters. The consistency of these two $\rho^{HX}_{00,in}$ confirms the validity of our method. From this figure, $\rho^{HX}_{00,in}$ is less than 1/3 when $\Phi \in[0,\pi/8)$, and $\rho^{HX}_{00,in}$ is greater than 1/3 when $\Phi \in[3\pi/8,\pi/2)$, which is consistent with our prediction as shown in Fig.~\ref{fig:3}.

Fig.~\ref{fig:7} shows the comparison of corrected $\rho_{00,real}$, observed $\rho_{00}$ from event sample ``Data" and the input $\rho_{00,input}$ as function of $\Phi-\Psi$. While the observed $\rho_{00}$ show clearly deviation from $\rho_{00,input}$, the corrected $\rho_{00,real}$ are consistent with the input $\rho_{00,input}$ within uncertainties for all $\Phi-\Psi$ bins.

The above results from the Monte-Carlo simulation verified the feasibility of our method to correct the effects of finite acceptance on the measurement of the azimuthal dependence of $\phi$ mesons global spin alignment. The method we proposed can be used in experiments to correct for effects arising from finite coverage of detectors when measuring the azimuthal dependence of the global spin alignment.

\section{\label{sec:sum}Summary}

The measurement of the azimuthal dependence of global spin alignment of vector mesons provide insight into the vorticity generated in the produced hot and dense matter. In this work, we propose a method to correct for the effects of finite pseudo-rapidity coverage on the azimuthal dependence of extracted spin alignment parameters $\rho_{00}$. The effect of finite $\eta$ cut on the  $\phi$ meson spin alignment $\rho_{00,acc}$ and the intrinsic spin alignment $\rho^{HX}_{00,in}$ are both taken into consideration. The method we proposed has been examined by a toy model simulation, The procedure we laid out in this paper will help us to measure the azimuthal dependence of vector meson global spin alignment in experiments.

\begin{acknowledgments}
We would like to thank A.~H.~Tang, C.~S.~Zhou, L.~Zheng for fruitful discussions. C.Z. acknowledges the support from the Chinese Scholarship Council. This work is supported in part by National Natural Science Foundation of China under Grants No.11890711 (11890710).
\end{acknowledgments}


\begin{thebibliography}{999}
\bibitem{Liang:2004ph}
Z.~T.~Liang and X.~N.~Wang,
Phys. Rev. Lett. \textbf{94}, 102301 (2005)
[erratum: Phys. Rev. Lett. \textbf{96}, 039901 (2006)]

\bibitem{Liang:2004xn}
Z.~T.~Liang and X.~N.~Wang,
Phys. Lett. B \textbf{629}, 20-26 (2005)

\bibitem{Betz:2007kg}
B.~Betz, M.~Gyulassy and G.~Torrieri,
Phys. Rev. C \textbf{76}, 044901 (2007)

\bibitem{Gao:2007bc}
J.~H.~Gao, S.~W.~Chen, W.~t.~Deng, Z.~T.~Liang, Q.~Wang and X.~N.~Wang,
Phys. Rev. C \textbf{77}, 044902 (2008)

\bibitem{STAR:2017ckg}
L.~Adamczyk \textit{et al.} [STAR],
Nature \textbf{548}, 62-65 (2017)

\bibitem{Becattini:2007sr}
F.~Becattini, F.~Piccinini and J.~Rizzo,
Phys. Rev. C \textbf{77}, 024906 (2008)

\bibitem{Pang:2016igs}
L.~G.~Pang, H.~Petersen, Q.~Wang and X.~N.~Wang,
Phys. Rev. Lett. \textbf{117}, no.19, 192301 (2016)

\bibitem{Jiang:2016woz}
Y.~Jiang, Z.~W.~Lin and J.~Liao,
Phys. Rev. C \textbf{94}, no.4, 044910 (2016)
[erratum: Phys. Rev. C \textbf{95}, no.4, 049904 (2017)]

\bibitem{Becattini:2015ska}
F.~Becattini, G.~Inghirami, V.~Rolando, A.~Beraudo, L.~Del Zanna, A.~De Pace, M.~Nardi, G.~Pagliara and V.~Chandra,
Eur. Phys. J. C \textbf{75}, no.9, 406 (2015)
[erratum: Eur. Phys. J. C \textbf{78}, no.5, 354 (2018)]

\bibitem{Zhang:2019xya}
J.~j.~Zhang, R.~h.~Fang, Q.~Wang and X.~N.~Wang,
Phys. Rev. C \textbf{100}, no.6, 064904 (2019)

\bibitem{Huang:2011ru}
X.~G.~Huang, P.~Huovinen and X.~N.~Wang,
Phys. Rev. C \textbf{84}, 054910 (2011)

\bibitem{Florkowski:2019qdp}
W.~Florkowski, A.~Kumar, R.~Ryblewski and R.~Singh,
Phys. Rev. C \textbf{99}, no.4, 044910 (2019)

\bibitem{Xia:2020tyd}
X.~L.~Xia, H.~Li, X.~G.~Huang and H.~Zhong Huang,
Phys. Lett. B \textbf{817}, 136325 (2021)

\bibitem{Li:2017slc}
H.~Li, L.~G.~Pang, Q.~Wang and X.~L.~Xia,
Phys. Rev. C \textbf{96}, no.5, 054908 (2017)

\bibitem{ALICE:2019aid}
S.~Acharya \textit{et al.} [ALICE],
Phys. Rev. Lett. \textbf{125}, no.1, 012301 (2020)

\bibitem{ALICE:2019onw}
S.~Acharya \textit{et al.} [ALICE],
Phys. Rev. C \textbf{101}, no.4, 044611 (2020)

\bibitem{Singh:2018uad}
R.~Singh [ALICE],
Nucl. Phys. A \textbf{982}, 515-518 (2019)
\bibitem{Singha:2020qns}
S.~Singha [STAR],
Nucl. Phys. A \textbf{1005}, 121733 (2021)
\bibitem{STAR:2018gyt}
J.~Adam \textit{et al.} [STAR],
Phys. Rev. C \textbf{98}, 014910 (2018)

\bibitem{STAR:2008lcm}
B.~I.~Abelev \textit{et al.} [STAR],
Phys. Rev. C \textbf{77}, 061902 (2008)

\bibitem{Schilling:1969um}
K.~Schilling, P.~Seyboth and G.~E.~Wolf,
Nucl. Phys. B \textbf{15}, 397-412 (1970)
[erratum: Nucl. Phys. B \textbf{18}, 332 (1970)]


\bibitem{Fu:2021pok}
B.~Fu, S.~Y.~F.~Liu, L.~Pang, H.~Song and Y.~Yin,
Phys. Rev. Lett. \textbf{127}, no.14, 142301 (2021)

\bibitem{Sun:2021nsg}
Y.~Sun, Z.~Zhang, C.~M.~Ko and W.~Zhao,
[arXiv:2112.14410 [nucl-th]].

\bibitem{Becattini:2021iol}
F.~Becattini, M.~Buzzegoli, G.~Inghirami, I.~Karpenko and A.~Palermo,
Phys. Rev. Lett. \textbf{127}, no.27, 272302 (2021)

\bibitem{Niida:2018hfw}
T.~Niida [STAR],
Nucl. Phys. A \textbf{982}, 511-514 (2019)

\bibitem{Tang:2018qtu}
A.~H.~Tang, B.~Tu and C.~S.~Zhou,
Phys. Rev. C \textbf{98}, no.4, 044907 (2018)

\bibitem{Lan:2017nye}
S.~Lan, Z.~W.~Lin, S.~Shi and X.~Sun,
Phys. Lett. B \textbf{780}, 319-324 (2018)

\bibitem{Sjostrand:2014zea}
T.~Sj\"ostrand, S.~Ask, J.~R.~Christiansen, R.~Corke, N.~Desai, P.~Ilten, S.~Mrenna, S.~Prestel, C.~O.~Rasmussen and P.~Z.~Skands,
Comput. Phys. Commun. \textbf{191}, 159-177 (2015)

\bibitem{STAR:2007ccu}
B.~I.~Abelev \textit{et al.} [STAR],
Phys. Rev. C \textbf{76}, 024915 (2007)
[erratum: Phys. Rev. C \textbf{95}, no.3, 039906 (2017)]

\bibitem{STAR:2007mum}
B.~I.~Abelev \textit{et al.} [STAR],
Phys. Rev. Lett. \textbf{99}, 112301 (2007)
\end{thebibliography}
\end{document}